\documentclass[%
 reprint,
 amsmath,amssymb,
 aps,
]{revtex4-1}

\usepackage{graphicx}
\usepackage{amsmath}

\usepackage[squaren]{SIunits}
\newcommand{\muon}{\micro}

\newcommand{\Pt}{\ensuremath{p_{\text{T}}}}
\newcommand{\Meff}{\ensuremath{M_{\text{eff}}}}
\newcommand{\Minv}{\ensuremath{M_{\text{inv}}}}
\newcommand{\MET}{\ensuremath{E_{\text{T}}^{\text{miss}}}}

\newcommand{\Top}{\ensuremath{\text{t}}}
\newcommand{\Beauty}{\ensuremath{\text{b}}}
\newcommand{\sTop}{\ensuremath{\tilde{\text{t}}_{1}}}
\newcommand{\tTbar}{\ensuremath{\Top\,\bar{\Top}}}

\usepackage{color}

\begin{document}

\preprint{PRD/???}

\title{Reconstruction of stop quark mass at the LHC}

\author{Diego Casadei}

\author{Rostislav Konoplich}
\altaffiliation[Also at ]{%
  Manhattan College, Riverdale, New York, NY 10471
}
\affiliation{%
  Department of Physics, New York University,\\
  4 Washington Place, New York, NY 10003
}%

\author{Rashid Djilkibaev}
\affiliation{%
  Institute for Nuclear Research, Moscow, Russia 117312
}

\date{\today}

\begin{abstract}
The cascade mass reconstruction approach was applied to simulated 
production of the lightest stop quark at the LHC in the
cascade decay $\tilde{g} \to \sTop \, \Top \to \tilde{\chi}_{2}^{0}
\, \Top \, \Top \to \tilde{\ell}_{R} \, \ell \, \Top \, \Top \to
\tilde{\chi}_{1}^{0} \, \ell \, \ell \, \Top \, \Top$ with top quarks
decaying into hadrons.  The stop quark mass was reconstructed assuming
that the masses of gluino, slepton and of the two lightest neutralinos
were reconstructed in advance.

A data sample set for the SU3 model point containing 400k SUSY events
was generated which corresponded to an integrated luminosity of about
20 $\rm fb^{-1}$ at 14 TeV.  These events were passed through the
AcerDET detector simulator, which parametrized the response of a
generic LHC detector.  The mass of the $\tilde{t}_{1}$ was
reconstructed with a precision of about $10\%$.
\end{abstract}

 \pacs{Valid PACS appear here}
\maketitle


\section{Introduction}\label{sec-intro}

If supersymmetry exists at an energy scale of ~1 TeV, the study of
third generation sleptons and squarks at the LHC is of a special
interest.  Their masses can be very different from that of sparticles
of the first and second generation, because of the effects of large
Yukawa and soft couplings as can be seen from the renormalization group
equations. Furthermore they can show large mixing in pairs
$(\tilde{t}_{L}, \tilde{t}_{R}), (\tilde{b}_{L}, \tilde{b}_{R})$ and
$(\tilde{\tau}_{L}, \tilde{\tau}_{R})$.  A detailed discussion of
possible SUSY effects at the LHC is given in \cite{atlas}.

\begin{figure}[b]
  \centering
  \includegraphics[width=0.9\columnwidth]{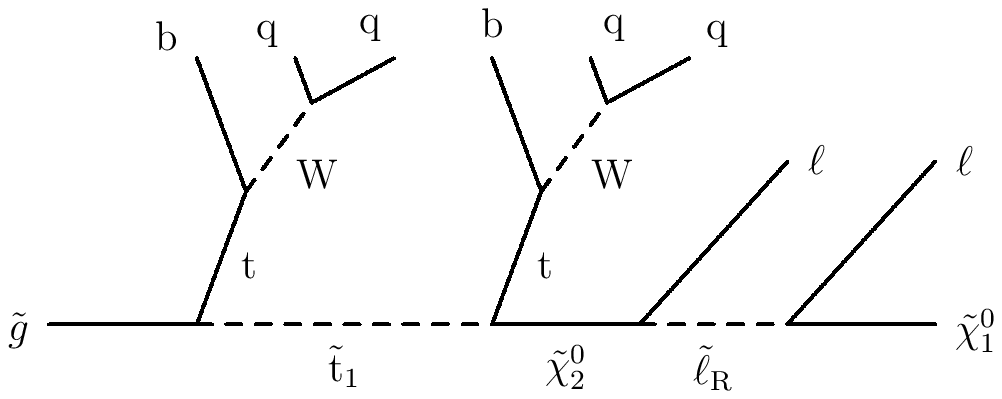}
  \caption{A gluino cascade decay chain with a stop quark production.}
\label{fig-chain}
\end{figure}

In this paper we consider the mass reconstruction of the lightest stop
quark ($\tilde{t}_{1}$) in the cascade decay
\begin{equation}
\tilde{g} \to \sTop \, \Top \to \tilde{\chi}_{2}^{0} \, \Top \, \Top
          \to \tilde{\ell}_{R} \, \ell \, \Top \, \Top
          \to \tilde{\chi}_{1}^{0} \, \ell \, \ell \, \Top \, \Top 
\label{eq-chain} 
\end{equation}
with the top quarks decaying into hadrons.  The gluino decay chain
(\ref{eq-chain}) is represented in Fig.~\ref{fig-chain}, in which
all final state particles are explicitly shown.  Here, the considered
leptons are electrons and muons ($\ell=$ e, \muon).  The lightest
neutralino $\tilde{\chi}_{1}^{0}$ is invisible to the particle
detector, whereas b quarks and light quarks (labeled as `q' in
Fig.~\ref{fig-chain}) are observed as jets.

Approaches to stop quark mass reconstruction in different decay
chains and for different points in the MSSM parameter space are
discussed in literature (for example in \cite{gj_note,noj1,noj2,krstic,raklev,mehdi}).
Recent limits from searches for stop quarks were published in
\cite{cdf}.

The reconstruction of a SUSY event is complicated because of the
escaping neutralinos and of the many complex and competing decay
modes.  At present, there are two different approaches to SUSY mass
reconstruction.  The \emph{endpoint method}, which has been widely
studied
\cite{gj_note,noj1,noj2,baer,cms,endpoints,allan,gjelsten0,gjelsten,lester,bar1,bar2},
looks for kinematic endpoints of invariant mass distributions.  The
second method is the \emph{mass relation approach}
\cite{tovey,nojiri,dk,stau}, based on the ``mass relation equation''
which relates the masses of the SUSY particles and the measured
momenta of the detected particles. It was shown in \cite{dk} that the
mass relation approach can be successfully used for integrated
luminosities as low as a few $\rm fb^{-1}$.

In this work, the mass relation approach of \cite{dk} is used for
measuring the mass of the lightest stop quark $\sTop$, assuming
about 20 $\rm fb^{-1}$ of integrated luminosity in LHC proton-proton
collisions at $\sqrt{s}=14$ TeV, under the assumption that the masses
of the gluino, slepton and of the two lightest neutralinos have been
reconstructed in advance with 10--20\% uncertainty.  At such
low integrated luminosity, the stop mass reconstruction is quite
challenging because of a high level of SUSY background (sparticles created
in decay chains different from that of Fig.~\ref{fig-chain}) and Standard
Model $\tTbar$ background.

In this paper, a particular example is chosen (the SU3 model;
section~\ref{sec-simul}) to illustrate the method.  First, an ``event
filter'' is applied to suppress the background
(section~\ref{sec-bkg-supp}), making use of a likelihood function
built with the known uncertainties on jet and lepton measurements, constrained
by the mass relation equation.  Next, events which do not satisfy the
kinematic limits derived from the chain (\ref{eq-chain}) are discarded
(section~\ref{sec-limits}).  Finally, a combinatorial mass
reconstruction method (section~\ref{sec-stop-mass}) is applied to find
the best estimate of the stop mass from the maximization of a combined
likelihood function, which depends on all five sparticle masses
(gluino, stop, slepton and the two lightest neutralinos) and is
constructed for each possible permutation of the final state
particles.

\section{Simulation}\label{sec-simul}

The particular mSUGRA model chosen for this work is the bulk point
SU3, one of the official benchmark points of the ATLAS collaboration
\cite{atlas}, which is compatible with the recent precision WMAP data
\cite{wmap}.  The values of the relevant mSUGRA parameters are given
in Table~\ref{tab-paramth}.  For the SUSY particles in the cascade
process~(\ref{eq-chain}), the theoretical masses and the production
cross section have been found by ISAJET 7.74 \cite{isajet} as reported
in Table~\ref{tab-massth}.

\begin{table}[t]
\begin{ruledtabular}
 \begin{tabular}{cccccc}
   Point & $m_{0}$ & $m_{1/2}$ & $A_{0}$ & $\tan\beta$ & $\mu$ \\
         & (GeV)   & (GeV)     & (GeV)   &            &       \\
   \hline
   SU3   & 100     & 300       & $-300$  & 6          &  $>0$ \\
 \end{tabular}
\end{ruledtabular}
\caption{mSUGRA parameters for the SU3 point \cite{atlas}.}\label{tab-paramth}
\end{table}

The branching ratio for the gluino decay chain (\ref{eq-chain}) at the
SU3 point is
\begin{equation}
\nonumber
\tilde{g} \stackrel{25.2 \%}{\longrightarrow} \sTop
    \stackrel{11.5 \%}{\longrightarrow} \tilde{\chi}_{2}^{0} 
    \stackrel{11.4 \%}{\longrightarrow} \tilde{\ell}_{R}
    \stackrel{100 \%}{\longrightarrow} \tilde{\chi}_{1}^{0}
    \quad \Rightarrow \quad 0.33 \% . 
\end{equation}
Monte Carlo simulations of SUSY production for the SU3 model point
were performed with the HERWIG 6.510 event generator \cite{herwig}.
Later, the produced events were passed through the AcerDET detector
simulation \cite{atlfast}, which parametrized the response of a
generic LHC detector (ATLAS and CMS detector descriptions can be found
in \cite{det_atlas} and \cite{det_cms}).  The efficiency for b-jet
reconstruction and labeling was set to $80\%$, whereas the calorimeter
response to electrons and jets was
\begin{eqnarray}
  &\text{e} : \quad &\dfrac{\sigma}{E}
     = \dfrac{0.12}{\sqrt{E/\text{GeV}}} \; \oplus \; 0.005
     \label{eq-ele-res}
\\
  &\text{j} : \quad &\dfrac{\sigma}{E}
     = \dfrac{0.5}{\sqrt{E/\text{GeV}}} \; \oplus \; 0.03
     \label{eq-jet-res}
\end{eqnarray}
For muons, the same response function as for electrons has been used
as first approximation.

A sample of 400k SUSY events (including signal and background
processes) was generated.  This approximately corresponds to $20 ~\rm
fb^{-1}$ of integrated luminosity for the SUSY SU3 point production
cross section of 19 pb at 14 TeV.  The masses 
of $\tilde{g}$, $\tilde \chi_{2}^{0}$, $\tilde \ell_{R}$,
$\tilde \chi_{1}^{0}$ listed in
Table~\ref{tab-massth} were given as input and it was assumed that these
masses  had been already measured with about 10-20\%
uncertainty with the method explained in \cite{dk}.

In order to isolate the chain (\ref{eq-chain}) and to suppress the
background, the following selection cuts were applied to the
reconstructed quantities:
\begin{itemize}
  \item exactly two isolated opposite-sign same-flavor (OSSF) leptons
        (either electrons or muons) with transverse momentum $\Pt >
        20,10$ GeV;

  \item two b-tagged jets with $\Pt > 50$ GeV; 

  \item at least three jets with \Pt\ larger than 150, 100, and 50 GeV

  \item at least nine jets with $\Pt > 10$ GeV (including b-tagged
        jets);

  \item no $\tau$-tagged jets;

  \item $\Meff > 600$ GeV and $\MET > 0.2 \Meff$, where \Meff\ is 
        the scalar sum of the
        missing transverse energy and the transverse momenta of the
        four hardest jets and \MET\ is the
        missing transverse energy;

  \item lepton invariant mass $50~\text{GeV} < M_{\ell\ell} < 105$ GeV. 
\end{itemize}
A total of 24 signal and 191 background events are left after applying 
these cuts: the SUSY background to the signal
process (\ref{eq-chain}) is thus significant (the classification of events
as signal and background is based on the knowledge of the
simulated information).

\begin{table}[t]
\begin{ruledtabular}
 \begin{tabular}{ccccccc}
   Point & $m_{\tilde{g}}$ & $m_{\sTop}$ & $m_{\tilde{\chi}_{2}^{0}}$
     & $m_{\tilde{\ell}_{R}}$ & $m_{\tilde{\chi}_{1}^{0}}$ & $\sigma$ \\
       & (GeV)  & (GeV)  & (GeV)  & (GeV)  & (GeV)  & (pb) \\
   \hline
   SU3 & 720.16 & 440.26 & 223.27 & 151.46 & 118.83 & 19 \\
 \end{tabular}
\end{ruledtabular}
\caption{Theoretical masses and total production cross section
$\sigma$ of SUSY particles at the SU3 point}\label{tab-massth}
\end{table}

As shown in \cite{gj_note}, the SM processes are suppressed
significantly by the above requirements. The SM dominant background
surviving the hard cuts is $\tTbar$ production, where both W
bosons decay leptonically producing a $\Beauty\,\Beauty\,l\,l$ state.  Since the
$\tTbar$ production cross section is about 833 pb, a 17M
$\tTbar$ sample, corresponding to 20 fb$^{-1}$ of integrated
luminosity, was generated with the HERWIG event generator.  After
applying the above selection cuts, only 21 $\tTbar$ background
events survive.

Note that the requirement of high hadronic activity is important for
the $\tTbar$ background suppression.  If the cut on the total
number of jets $N_{\text{jet}} \ge 9$ is loosened to 7 jets, the number
of $\tTbar$ events surviving the selection cuts increases to 115.

\begin{figure}
  \centering
  \includegraphics[width=0.9\columnwidth]{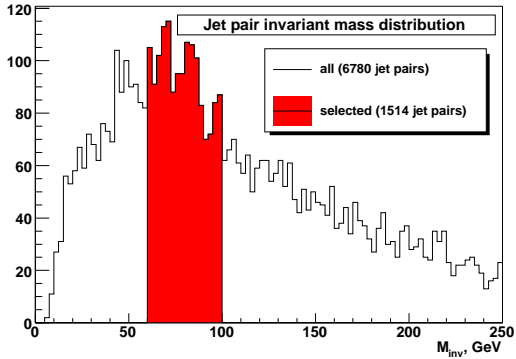} 
  \caption{Light jet pair invariant mass distribution.  Only j\,j pairs
  with mass in the range 60--100 GeV have been retained in the analysis.}
\label{fig-mW}
\end{figure}

For every event, light jets (i.e.\ not tagged as b-jets) were combined
in pairs whose invariant mass has been reconstructed.  Only the
independent pairs whose \Minv\ is in the range 60--100 GeV (W boson
region) have been retained and events with less than two jet pairs
have been rejected.  The invariant mass distribution obtained with all
combinations is shown in Fig.~\ref{fig-mW}.

\begin{figure}
  \centering
  \includegraphics[width=0.9\columnwidth]{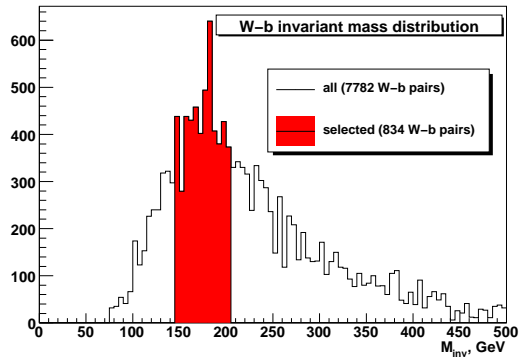} 
  \caption{W-b invariant mass distribution.  W-b pairs are selected in
  the range from 145 GeV to 205 GeV.}
\label{fig-mtop}
\end{figure}

Next, all combinations of a j\,j pair and a b-jet have been
considered, retaining only those with $\Minv=$145--205 GeV,
representing the acceptable W-b pairs in the top quark region
(Fig.~\ref{fig-mtop}).  After this step, from the initial 215 SUSY
events (signal + background), a total of 834 b\,j\,j combinations were
identified as candidates for a top quark decay $\Top\rightarrow$ b\,W
$\rightarrow$ b\,j\,j, coming from 23 signal and 70 background events.
Thus at this stage only one signal event was lost.  On the other hand,
applying the same procedure to the 21 $\tTbar$ background events left
only 3 events with 21 b\,j\,j combinations as candidates for a top
quark decay.

Naively, one would think that this selection should let all
$\tTbar$ background events survive, but it is clearly not the case.
If a W-boson decays leptonically an invariant mass of light jets is not
related with the W-boson mass. Also a parton showering leads to 
redistribution of energy and this means that the
momenta of the b-quark and the W-boson are not sufficient to give the
top invariant mass.  The final result is that most $\tTbar$ events
are dropped at this stage.


\section{Background suppression}\label{sec-bkg-supp}

To suppress the background before the last (combinatorial) step, an
event filter is used which assumes that the masses of $\tilde
\chi_{2}^{0}$, $\tilde \ell_{R}$, $\tilde \chi_{1}^{0}$ are known
(Table~\ref{tab-massth}).  The event filter maximizes, for each event,
a likelihood function constrained by the mass relation equation, or
equivalently minimizes the function:
\begin{equation}
\chi^2(m_{\tilde{g}}, m_{\tilde{b}}) =
   \sum_{i=1}^4 \frac{(p_i^{\text{event}}-p_i^{\text{meas}})^2}{\sigma_{i}^2}
  + \lambda f(\vec{m}, \vec{p})
\label{chi1}
\end{equation}
where the index $i$ runs over the two leptons and the two t-quarks,
$p_{i}^{\text{meas}}$ and $\sigma_i$ are the reconstructed momentum
and its uncertainty, and $p_{i}^{\text{event}}$ is the true momentum,
which is varied to find the minimum
(only uncertainties in jet and lepton energy measurements are taken
into account).  The parameter $\lambda$ is a Lagrange multiplier and
$f(\vec{m}, \vec{p})=0$ is the mass relation.

The constraint $f(\vec{m}, \vec{p})=0$ is the key of the ``mass
relation approach'': it relates the masses of the SUSY particles to
the measured momenta of the final state particles in the chain
(\ref{eq-chain}) \cite{tovey,nojiri,dk}.  The mass relation constraint
is obtained as a solution of a system of four-momentum constraints for
each vertex containing SUSY particles in the decay chain
(\ref{eq-chain}).  For example, for the gluino decay vertex one has
$m_{\tilde{g} }^2 = (p_{\tilde{\chi}_{1}^{0}} 
                  + k_{l_{1}} + k_{l_{2}} 
                  + p_{t_{1}} + p_{t_{2}})^{2}
$
where the right hand part is in terms of the four-momenta 
of the lightest supersymmetric particle
(LSP), the detectable leptons, and the reconstructed top quarks.
Similar relations can be obtained for each SUSY vertex in the process
(\ref{eq-chain}), which contains four vertices, hence one gets four
kinematic equations that can be solved to find the four components of
$p_{\tilde{\chi}_{1}^{0}}$ in terms of the SUSY masses and the momenta
of the detectable and reconstructed particles.  By substituting these
components into the on-shell mass condition for the LSP, $m_{\tilde
\chi_{1}^{0}}^{2} = p_{\tilde \chi_{1}^{0}}^{2}$, the mass relation
constraint includes all SUSY masses ($\vec{m}$) and the momenta
($\vec{p}$) of the detectable leptons and of the reconstructed
t-quarks in the process (\ref{eq-chain}).  The explicit form of
$f(\vec{m}, \vec{p}) = 0$ is given by \cite{tovey,nojiri,dk}.

Note that in the decay chain (\ref{eq-chain}) the locations of each 
of the two t-quarks and of each of the
two leptons  are unknown.  Here, it
is assumed that the t-quark with higher energy originates from the
gluino decay.  In addition, we assumed that the leptons with higher
$p_T$ originate from the $\tilde\chi_2$ decay.  The momentum
resolution for a t-quark is computed as $\sigma =
\sigma_{\text{b-jet}} \,\oplus\, \sigma_{\text{jet 1}} \,\oplus\,
\sigma_{\text{jet 2}}$, where all jets are assumed to have the
resolution mentioned above (equation~(\ref{eq-jet-res}) in
section~\ref{sec-simul}).

In the numerical minimization procedure by MC sampling, the gluino
mass is left free to vary within $\pm 20\%$ from the value reported in
Table~\ref{tab-massth} (a Gaussian sampling), and the stop mass is
left free to vary uniformly in the range 480$\pm$120 GeV because the
stop mass is unknown at this step.  The latter range has lower limit
approximately related with the kinematic condition for a stop decay
$m_{\sTop} > m_{\tilde{\chi}_{2}^{0}} + m_{\Top}$ and is
limited from above by the sbottom quark mass, which is about 600 GeV
for the SU3 point.

\begin{figure}
  \centering
  \includegraphics[width=0.9\columnwidth]{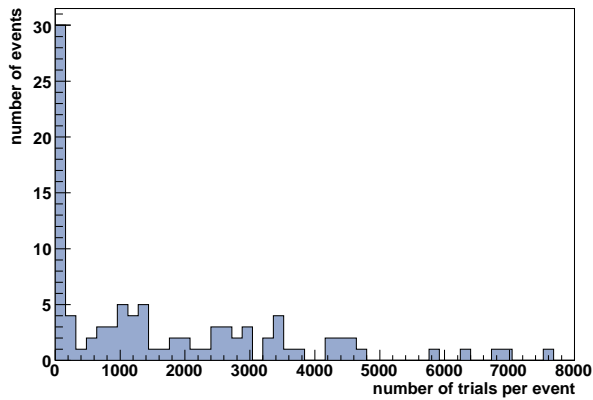}
  \caption{The number of events versus the number of accepted trials
  per event with $\chi^2 < 10$. The total number of trials is $10^5$.}
\label{fig-evfilter}
\end{figure}

For signal events, the event likelihood distribution has a maximum in
the region of the ($\tilde{g}$, $\sTop$) mass plane correlated
with the true masses of $\tilde{g}$ and $\sTop$.  Hence, signal
events should give a peak in the region of the true masses.  On the
other hand, for background events there is no strong correlation
between the likelihood maximum and the true masses of $\tilde{g}$ and
$\sTop$.  Therefore, if we arbitrarily chose a point in the
($\tilde{g}$, $\sTop$) mass plane in the range close to true
masses, there is a very high probability that the value of $\chi^2$
found from equation~(\ref{chi1}) is smaller for a signal event than
for a background one.  For each event, $10^5$ points in the mass plane
were generated in the range $720 \pm 144$ GeV for the gluino
and $480 \pm 120$ GeV for the stop quark, and the $\chi^2$ was
calculated. Results of event filter are presented in 
Fig.~\ref{fig-evfilter}. This figure shows the number of events versus 
the number of trials per event in which $\chi^2 < 10$. Two groups of
events can be seen: a group with the number of accepted trials close
to zero and a group with more than a few hundred of accepted trials.
The first group presents background events for which unlike for signal 
events a minimum of $\chi^2$ need not be in the considered mass range.
Thus if $\chi^2 < 10$ in at least 300 trials per event, this event 
was considered as a signal
candidate and was retained for the subsequent analysis.

Before the application of the event filter, we had 23 signal events,
70 SUSY background events and 3 Standard Model $\tTbar$ background
events.  After the application of the event filter, the SUSY
background events were reduced approximately by a factor of 2 while a
single signal event was lost: 22 signal events, 37 SUSY
background events and 3 $\tTbar$ background events survived.

\section{Kinematic limits}\label{sec-limits}

It was shown
\cite{gj_note,noj1,noj2,baer,cms,endpoints,allan,gjelsten0,gjelsten,lester,bar1,bar2}
that the endpoint method could be very useful in SUSY particle mass
reconstruction for finding relations between the masses of the SUSY
particles involved in a decay chain and to determine their masses.
Such a method allows mass reconstruction without relying on a specific
SUSY model.  In particular, the endpoint method can be applied to a
decay chain of the type
\begin{equation}
A \to b \, B \to b \, c \, C .
\label{eq2} 
\end{equation}
where particles $A$, $B$, $C$ are invisible but particles $b$ and $c$
are considered as visible (they can be either directly detected or
indirectly reconstructed from the properties of final state
particles).

In the literature, kinematic limits on the invariant mass distribution
of $b\,c$ pairs in decay (\ref{eq2}) over a variable $q^2=(p_b+p_c)^2$
often are given for the case in which at least one of the visible
particles is massless.  However, in some cases both particles $b$ and
$c$ can have a non-negligible mass.  For example, this is the case when
a gluino decays into a stop quark and top quark.  These kinematic
limits for the case of all massive particles in process (\ref{eq2})
are given \cite{lester_thes,bartl,student} by
\begin{equation}
  q = \sqrt{(-R \pm \sqrt{R^2-4QS})/2Q}
 \label{limits}
\end{equation}
with
\begin{eqnarray}
&\hspace{-3ex}&
\begin{array}{lll}
  Q & = & M_B^2
\end{array}
\\
&\hspace{-3ex}&
\begin{array}{lll}
  R  & = & (m_b^2-M_A^2-M_B^2) \, (m_b^2+m_c^2) \; +      \\
     &   & + \, (m_b^2-M_A^2+M_B^2) \, (M_B^2-m_b^2-M_C^2) \\
\end{array}
\\
&\hspace{-3ex}&
\begin{array}{llll}
  S  & = & M_A^2 \, (m_b^2-m_c^2)^2  & \!\! + \\
     &   & + \, (M_A^2-M_C^2) & \!\!\!\!\! [m_b^2 \, (M_B^2-m_b^2-M_C^2) \; + \\
     &   &                    & \!\!\! + \, m_c^2 \, (M_A^2+m_b^2-M_B^2)]
\end{array}
\end{eqnarray}
where the upper edge corresponds to the case when $b$ and $c$ particles
are moving in opposite directions in the rest frame of particle $A$
and the lower edge corresponds to the case when $b$ and $c$ particles
are moving in the same direction.  A nonzero lower limit is a
consequence of nonzero masses of the particles.

The kinematic limits from equation (\ref{limits}) for the process
(\ref{eq-chain}) are $q_{\text{min}}$ = 375.1 GeV and $q_{\text{max}}$
= 496.8 GeV for particles created on-shell.  Because
SUSY particles can also be created off mass shell, we have set a wider
range.  By choosing a cut at 525 GeV in $\tTbar$ invariant mass,
the number of events remaining after the event filter selection is
reduced to 21 signal events, 34 SUSY background events and 3
$\tTbar$ background events.

\section{Stop quark mass reconstruction}\label{sec-stop-mass}

If all sparticle masses but the stop mass in the decay chain (\ref{eq-chain})
were known, the mass relation equation would allow finding
the stop mass directly. However, we assume that these masses are known
at integrated luminosity of 20 $\rm fb^{-1}$ with an uncertainty of
10-20$\%$. In order to take this into account, at the final step of
the stop mass reconstruction we allow for all sparticle masses to vary
in ranges defined by their uncertainties.  Because in this case for
each event there are five unknown masses, at least five events are
required to reconstruct sparticle masses.

The combinatorial procedure of \cite{dk} is used for the final stop mass
reconstruction, applied only to the events that pass the event filter.
It is important to make the most effective cuts in advance, because
such procedure is computationally very intensive due to combinatorics: 
all possible groupings of five events in the sample are considered.

For each set of five events, a combined likelihood function is built
and maximized.  For sparticle masses the combined likelihood function
for the combination is defined as the product of the maximum
likelihood functions for individual events.  Finding a maximum of the
combined likelihood function for the combination is the same as 
searching for a minimum of the function

\begin{equation}
\chi^2_{\text{comb}}(\vec m)=\sum_{i=1}^5{\text{min} (\chi^2_{\;\text{event}})_i} .
\label{chi20}
\end{equation}
In Eq.~(\ref{chi20}), $\text{min} (\chi^2_{\;\text{event}})_i$ is a
result of searching for a minimum of the $\chi^2_{\;\text{event}}$
function for an individual event. For each of the five events in the
combination, the $\text{min} (\chi^2_{\;\text{event}})$ is fitted with
9 parameters (four particle momenta and five SUSY masses).

The $\chi^2$ function for an individual event is defined by
\begin{equation}
\begin{array}{ll}
  \chi^2_{\text{event}} & =
   \sum_{i=1}^4 \dfrac{(p_i^{\text{event}}-p_i^{\text{meas}})^2}{\sigma_{i}^2} \; + \\
  & + \sum_{n=1}^5 \dfrac{(m_n^{\text{event}}-m_n)^2}{\sigma_{n}^2}
  + \lambda_1 f + \lambda_2 f^{\ell\ell}
\end{array}
\label{chi2}
\end{equation}
where the variables with superscript ``event'' are those with respect
to which one has to minimize.

\begin{figure}
  \centering
  \includegraphics[width=0.9\columnwidth,height=0.6\columnwidth]{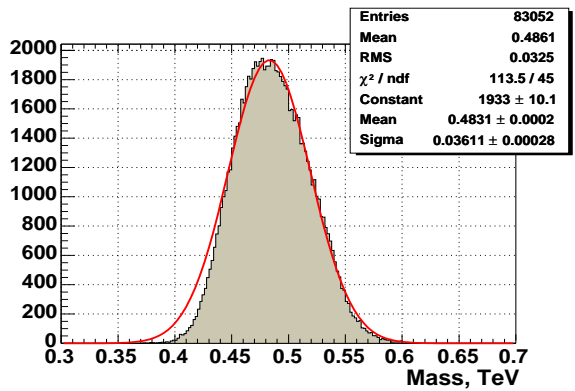} 
  \caption[Reconstructed stop mass.]{Reconstructed stop mass
  distribution including SUSY background and SM
  $\tTbar$ background with integrated luminosity of 20 $\rm
  fb^{-1}$.  The curve is the result of a Gaussian fit.}
\label{fig-stop}
\end{figure}

The first term in equation~(\ref{chi2}) takes into account the
deviations of the reconstructed momenta of the t-quarks, or the
measured momenta of the leptons, from the true ones. The second term
takes into account the intrinsic width of the SUSY particle masses, in
the Gaussian approximation (instead of a Breit-Wigner distribution).
The standard deviations corresponding to the mass widths are taken to
be 15 GeV for the gluino, 2 GeV for the stop and 1 GeV for all light
masses.  The first two numbers are comparable with the theoretical
widths for the heavy SUSY particles.  The last number takes into
account the fact that light SUSY particles are either quite narrow or
stable.  We note that the results of the mass reconstruction are not
strongly sensitive to the actual values of sparticle widths.  In
equation~(\ref{chi2}) the mass relation and the $\ell\,\ell$ edge,
which relates three light sparticle masses for chain (\ref{eq-chain}),
are included by means of the Lagrange multipliers $\lambda_{1}$ and
$\lambda_{2}$.  The $\ell\,\ell$ edge (103.1 GeV) can be obtained, for
example, from Eq.~(\ref{limits}) by plugging in the correspondent
light sparticle masses and zero lepton masses.

The minimization is done numerically by means of MC samplings of the
parameter space.  The sampling is uniform for the stop mass, in the
range: 480 $\pm $ 120 GeV.  For the masses of $\tilde{g}$,
$\tilde{\chi}_{2}^{0}$, $\tilde{\ell}_{R}$, $\tilde{\chi}_{1}^{0}$ a
Gaussian sampling is done, with mean values as given in the
Table~\ref{tab-massth} and standard deviations of 72, 20, 20, 20 GeV,
which approximately corresponds to uncertainties of $10\%, 10\%, 15\%,
20\%$ in sparticle masses found in \cite{dk}.  The MINUIT code
\cite{minuit} is used to search for the minimum of the
$\chi^2_{\text{comb}}$ function~(\ref{chi20}).

The reconstructed SUSY particle mass distribution, together with a
Gaussian fit, is shown in Fig.~\ref{fig-stop}.  As can be seen in
this figure, the reconstructed stop mass distribution is described
approximately by a Gaussian with the mean value of 483 GeV and 
standard deviation 36 GeV.

The asymmetry of the stop mass distribution toward higher masses can
be explained by the nearness of the kinematic edge for the stop decay
$\sTop \to \tilde{\chi}_{2}^{0} + \Top$ to
the stop mass so that high masses are generated more often than lower 
values.  This is not accounted for in the Monte Carlo sampling,
for which we have uniformly generated the values of the stop mass.

\begin{figure}
  \centering
  \includegraphics[width=0.9\columnwidth,height=0.6\columnwidth]{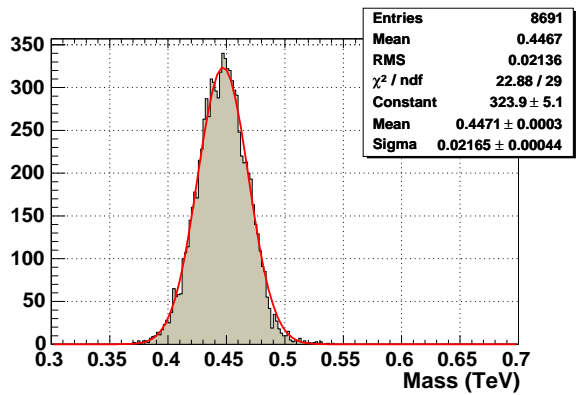} 
  \caption[Reconstructed stop mass for true signal
  events.]{Reconstructed stop mass distribution for true events
  without background with
  integrated luminosity of 20 $\rm fb^{-1}$.  The curve is the result
  of a Gaussian fit.}
\label{fig-idealstop}
\end{figure}

In order to understand the role of background and of the simulated
detector effects, the stop mass has been reconstructed using
generator-level momenta without SUSY background and Standard Model
$\tTbar$ background, and the result is shown in
Fig.~\ref{fig-idealstop}.  In this case, as expected, the
reconstructed stop mass is very close to the theoretical mass and the
width of the distribution is smaller because of the absence of
detector effects. It follows from comparison of Fig.~\ref{fig-stop}
and Fig.~\ref{fig-idealstop} that the presence of background tends
to shift the stop mass peak position to the region of higher masses.

\section{Conclusion}

We applied the cascade mass reconstruction approach developed in
\cite{dk} for reconstructing the mass of the 3$^{\text{rd}}$
generation supersymmetric quark, assuming 14 TeV proton-proton
collisions at the LHC with integrated luminosity of about $20 ~\rm
fb^{-1}$.  At such relatively low integrated luminosity, the stop mass
reconstruction is complicated because of a high level of SUSY and
Standard Model $\tTbar$ backgrounds and of the low branching ratio
for the gluino decay chain involving a stop quark.  Our approach to
the stop mass reconstruction is based on the consecutive use of an
event filter and of a combinatorial mass reconstruction method.

In this work, we considered the stop mass reconstruction at the SU3
mSUGRA point and we obtained an estimate of the stop mass with a
precision of about $10\%$.  We expect that our approach should work
for different MSSM parameters as well, provided that a decay chain
containing at least four successive two-body decays and involving five
SUSY particles has a sufficiently large branching ratio to be
identified in a heavy background environment.

\section*{Acknowledgments}  
The authors thank M.~Ibe, A.~Mincer and P.~Nemethy for interesting
discussions and useful suggestions.  This work has been supported by
the National Science Foundation under grant PHY-0854724.

\end{document}